\documentclass[british,12pt]{article}
\usepackage[utf8]{inputenc}
\usepackage{gensymb}

\usepackage[british]{babel}

\usepackage[margin=2.5cm,a4paper]{geometry} 
\usepackage[final]{hyperref} 

\usepackage[sorting=none,url=true,backend=bibtex,minbibnames=1,maxbibnames=3]{biblatex} 
\usepackage{csquotes}
\usepackage{tabularx} 
\usepackage{amsmath}  
\DeclareMathOperator{\arcsinh}{arcsinh}
\usepackage{amssymb}
\numberwithin{equation}{section}
\usepackage{mathtools}
\usepackage{siunitx}

\usepackage{graphicx,subcaption} 
\usepackage{wrapfig}
\usepackage{placeins}
\usepackage{float}
\usepackage{setspace}
\usepackage{etoolbox}
\usepackage{siunitx}
\BeforeBeginEnvironment{tcolorbox}{\savenotes}
\AfterEndEnvironment{tcolorbox}{\spewnotes}

\usepackage{cases}
\usepackage[T1]{fontenc}
\hypersetup{
	colorlinks=true,       
	linkcolor=blue,        
	citecolor=blue,        
	filecolor=magenta,     
	urlcolor=blue         
}

\usepackage[most]{tcolorbox}
\usepackage{empheq}

\usepackage{footnote}
\usepackage{bm}

\usepackage{duckuments}

\renewcommand{\lvert}{\left\vert}
\renewcommand{\rvert}{\right\vert}

\newcommand{\Ao}{\overline{A}}
\newcommand{\Bo}{\overline{B}}
\usepackage{perpage} 

\begin{document}

\begin{titlepage}

\vspace{5pt}

\begin{center}

{\Large\bf Velocity rotation curves in}\\
\vspace{10pt}

{\Large\bf the gravimagnetic dipole spacetime}\\

\vspace{10pt}

\vspace{40pt}

Cl\'ementine Dassy$^a$ and Jan Govaerts$^{a,b,c}$ 
\vspace{30pt}

$^{a}${\sl Centre for Cosmology, Particle Physics and Phenomenology (CP3),\\
Institut de Recherche en Math\'ematique et Physique (IRMP),\\
Universit\'e catholique de Louvain (UCLouvain),\\
2, Chemin du Cyclotron, B-1348 Louvain-la-Neuve, Belgium}\\
E-mail: {\em Clementine.Dassy@uclouvain.be}, {\em Jan.Govaerts@uclouvain.be}\\
ORCID: {\tt http://orcid.org/0000-0002-0965-7848}, {\tt http://orcid.org/0000-0002-8430-5180}\\
\vspace{15pt}
$^{b}${\sl International Chair in Mathematical Physics and Applications (ICMPA--UNESCO Chair)\\
University of Abomey-Calavi, 072 B.P. 50, Cotonou, Republic of Benin}\\
\vspace{15pt}
$^{c}${\sl Fellow of the Stellenbosch Institute for Advanced Study (STIAS),\\
Stellenbosch, Republic of South Africa}\\

\vspace{10pt}


\vspace{20pt}

\begin{abstract}
\noindent

The gravimagnetic dipole spacetime consists of two counter-rotating black holes of equal mass connected by a Misner string. For a particular distance in between them, the string is tensionless with the black holes at equilibrium with each other.

The geodesics of relativistic massive, or massless particles are considered, leading to the identification of circular rotation trajectories. The velocities of these trajectories are computed.

\end{abstract}

\end{center}

\end{titlepage}

\setcounter{footnote}{0}

\section{Introduction}

General relativity admits a wide range of exact solutions, many of which display non-trivial structures even in the absence of matter. Among these, the gravimagnetic dipole spacetime, a recently described \cite{MankoRodchenko, Clement2018, Clement2019, Clement2021} axisymmetric, asymptotically flat and stationary metric, consists of two counter-rotating NUT (Newman-Unti-Tamburino) black holes, of equal mass but opposite NUT charges, and connected by a Misner string. When the separation between the black holes is tuned precisely, the string becomes tensionless and the configuration is at equilibrium. By adimensionalising the system, the mass scale is removed and the gravimagnetic dipole spacetime is characterised by only one parameter, the NUT charge.

In this work, the motion of massive and massless test particles is studied in the equatorial plane, making full use of the available symmetries. Using a Hamiltonian formalism, an effective potential\cite{IgataT} is defined, leading to a condition for the existence of circular orbits. The rotation velocities associated with these orbits are then computed, thereby obtaining the velocity rotation curves for various values of the NUT parameter.

The recent study of the gravito-electromagnetic approximation to the gravimagnetic dipole \cite{Govaerts2023} derived an approximate velocity rotation curve for circular orbits within the weak-field regime of the gravimagnetic dipole spacetime, identifying conditions under which a roughly flat rotation curve emerges without invoking dark matter. The present exact results are compared with these approximate analytic expressions, in the relevant domain of parameters.

In the following, Section \ref{sec:gravimagneticdipole} reviews the gravimagnetic dipole metric. The conditions used to obtain a tensionless Misner string are described, allowing for the full characterisation of the spacetime with only one parameter, the NUT charge.

In Section \ref{sec:hamiltonianpotential}, the Hamiltonian for massive and massless particles in this spacetime is presented  and then used to define an effective potential for circular orbits. The number of circular orbits depends on the particle's energy and the value of the NUT parameter.

Section \ref{sec:distancevelocity} finally presents the calculation of the velocity rotation curves along with the corresponding figures. Comparison is made with the results from \cite{Govaerts2023}, showing good agreement in the relevant domain of parameters.

\section{The gravimagnetic dipole spacetime}\label{sec:gravimagneticdipole}

Let us consider the gravimagnetic dipole spacetime metric. This configuration \cite{MankoRodchenko,Clement2021} consists of the nonlinear superposition of two counter-rotating NUT objects of equal masses $m>0$ and opposite NUT charges $\pm \nu$ ($\nu\geq0$), separated by a total distance $2k \geq 2 m >0$, and positioned symmetrically on the $z$-axis relative to $z = 0$.

In Weyl coordinates $(x^0, x^i) = (ct,\rho, \phi, z)$, the metric can be written as
\begin{equation}\label{eq:metric}
ds^2 = -f (c dt - \omega d\phi)^2 + f^{-1} [e^{2 \gamma} (d \rho^2 + dz^2) + \rho^2 d \phi^2 ].
\end{equation}
The functions $f$, $\omega$ and $e^{2 \gamma}$ depend only on $(\rho,z)$ and are written as follows:
\begin{equation}
f = \frac{\lvert A \rvert^2 - \lvert B \rvert^2}{\lvert A + B \rvert^2}, \quad e^{2 \gamma} = \frac{\lvert A \rvert ^2 - \lvert B \rvert^2}{64 d^4 \alpha_+^2 \alpha_-^2 R_+ R_- r_+ r_-}, \quad \omega = - 4 \frac{\Im[G (\bar{A} + \bar{B})]}{\lvert A\rvert^2 - \lvert B \rvert^2},
\end{equation}
given the following definitions (with $m \equiv  G m /c^2$):
\begin{equation}\label{eq:definitions_notations}
\begin{aligned}
R_\pm(\rho,z) & = & \sqrt{\rho^2 + (z \pm \alpha_+)^2}, \qquad  & r_\pm(\rho,z) & =& \sqrt{\rho^2 + (z \pm \alpha_-)^2},\\
\alpha_\pm & = &  \sqrt{m^2 + k^2 - \nu^2 \pm 2 d}, \qquad  & d  & = & \sqrt{m^2 k^2 + \nu^2 (k^2 - m^2)}.
\end{aligned}
\end{equation}
Finally, the functions $A$, $B$ and $G$ (which also depend only on $(\rho,z)$) are given by the following expressions:
\begin{equation}
\begin{aligned}
A = & \frac{1}{2} \Big[ (d- m^2)^2 \alpha_+^2 + (d + m^2)^2 \alpha_-^2 \Big] (R_+ - R_-) (r_+ - r_-) \\
& - \alpha_+ \alpha_- \Big[2(d^2 -m^4)(R_+ R_- + r_+ r_-) + (d^2+ m^4)(R_+ +R_-)(r_++r_-) \Big] \\
& - 2 i m k \nu d \Big[ (\alpha_+ - \alpha_-)(R_+r_+ - R_- r_-) - (\alpha_+ + \alpha_-)(R_+r_- -R_-r_+)\Big], \\
B = &  - 4d \Big\{(d-m^2) \alpha_- [ m \alpha_+ (R_+ + R_-) + i k \nu (R_+ - R_-)] \\
&\qquad \quad  + (d + m^2) \alpha_+ [m \alpha_- (r_+ + r_-) + i k \nu(r_+ - r_-)] \Big\},
\end{aligned}
\end{equation}
\begin{equation}
\begin{aligned}
G = & d\big[ - (d-m^2)^2 \alpha_+ + (d +m^2)^2 \alpha_- - 2 i k \nu m^2 (\alpha_+ - \alpha_-)\big](R_+ r_+ - R_- r_-) \\
& + d\big[(d  - m^2)^2 \alpha_+ + (d+m^2)^2 \alpha_- + 2 i k \nu m^2 (\alpha_+ + \alpha_-)\big] (R_+ r_- - R_- r_+) \\
& - m(d^2 + m^4) \alpha_+ \alpha_- (R_+ +R_-)(r_+ + r_-) \\
&  + \frac{m}{2} \big[(d-m^2)^2 \alpha_+^2 + (d+m^2)^2 \alpha_-^2 + 8 i k \nu d^2\big] (R_+ - R_-)(r_+ - r_-) \\
& - 2 m (d^2 - m^4) \alpha_+ \alpha_- (R_+R_- + r_+ r_-)  \\
& - 2 d (d-m^2) \alpha_- \big[ (\alpha_+ + 2m - z)(m \alpha_+ + i k \nu) R_+ - \mathrlap{(\alpha_+ - 2m + z)(m \alpha_+ - i k \nu) R_-\big]} \\
&  - 2 d (d+m^2) \alpha_+ \big[ (\alpha_- + 2m - z)(m \alpha_- + i k \nu) r_+ - \mathrlap{(\alpha_- - 2m + z)(m \alpha_- - i k \nu) r_-\big]}.
\end{aligned}
\end{equation}

The functions $f$ and $e^{2 \gamma}$ are dimensionless; $\omega$ has a dimension of length, ensuring that every term in eq.~\eqref{eq:metric} has proper dimensions.

For ease of use, here are the non-vanishing entries of the metric and its inverse:
\begin{equation}
\begin{aligned}
&g_{tt}  = - f,&  g_{t \phi} & = f \omega, & g_{\phi \phi} = \frac{\rho^2}{f} - f \omega^2, & \qquad g_{\rho\rho} = g_{zz} = f^{-1} e^{2 \gamma}; & \\
&g^{tt}  = - f^{-1} + \frac{f \omega^2}{\rho^2}, &  g^{t \phi}  & = \frac{f \omega}{\rho^2}, &  g^{\phi \phi} = \frac{f}{\rho^2}, & \qquad   g^{\rho \rho} = g^{zz} = f e^{-2 \gamma} .&
\end{aligned}
\end{equation}
This metric is asymptotically flat, with total mass $M = 2 m$ and total angular momentum $J = 2 k \nu$.

\subsection{The values of $m$, $k$ and $\nu$}

\begin{figure}
\centering
\includegraphics[scale=0.5]{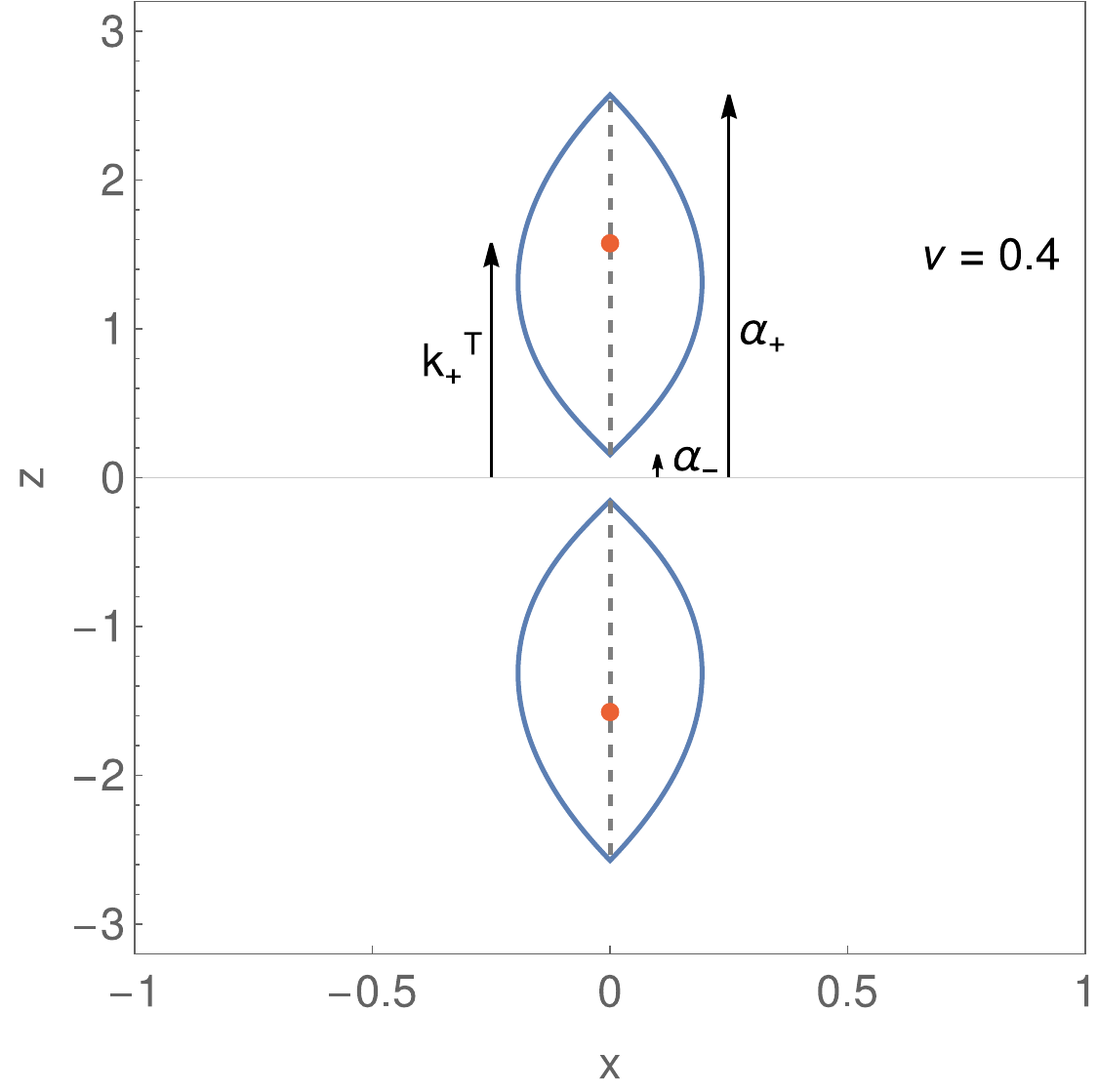}
\caption{\label{fig:horizon_details}This graph shows a cut in the XZ plane of the spacetime. The plain blue lines represent the static limit and the orange dots the black holes. The dashed grey lines represent the horizons. Various quantities ($\alpha_\pm$, the limits of the horizons on the vertical axis) as well as the separation of the black holes $2k^T_+$ are also shown, for a gravimagnetic dipole spacetime with NUT parameter $\nu = 0.4$.}
\end{figure}

\begin{figure}
\begin{minipage}[c]{0.49\linewidth}
\centering
    \includegraphics[width=\textwidth]{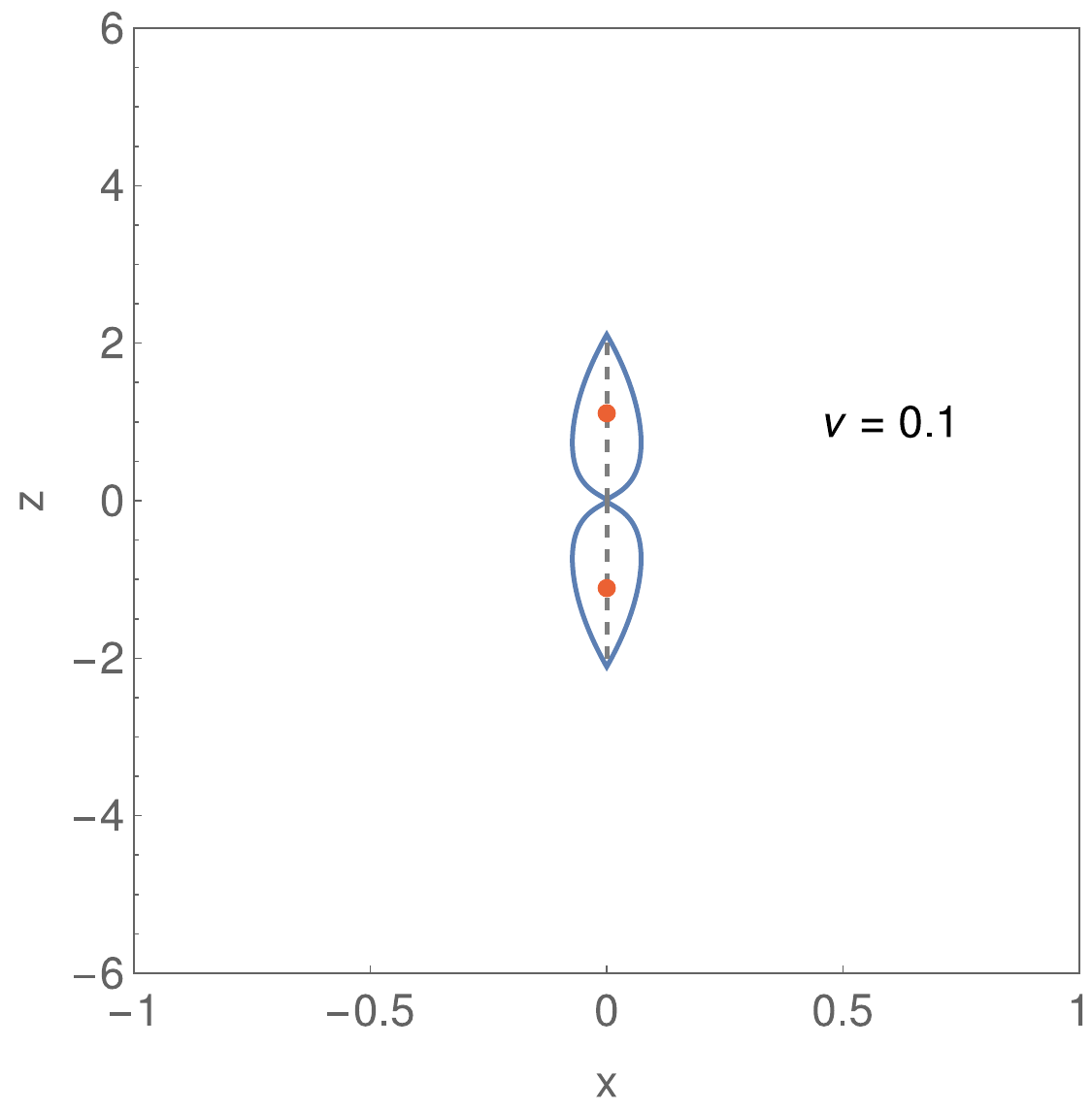}
    \caption{Cut of the spacetime along the $XZ$ plane for $\nu = 0.1$. The orange dots represent the black holes. The blue lines represent the static limit. The dashed grey lines represent the horizon rods.}
    \label{fig:horizon1}
\end{minipage}
\hfill
\begin{minipage}[c]{0.49\linewidth}
\centering
    \includegraphics[width=\textwidth]{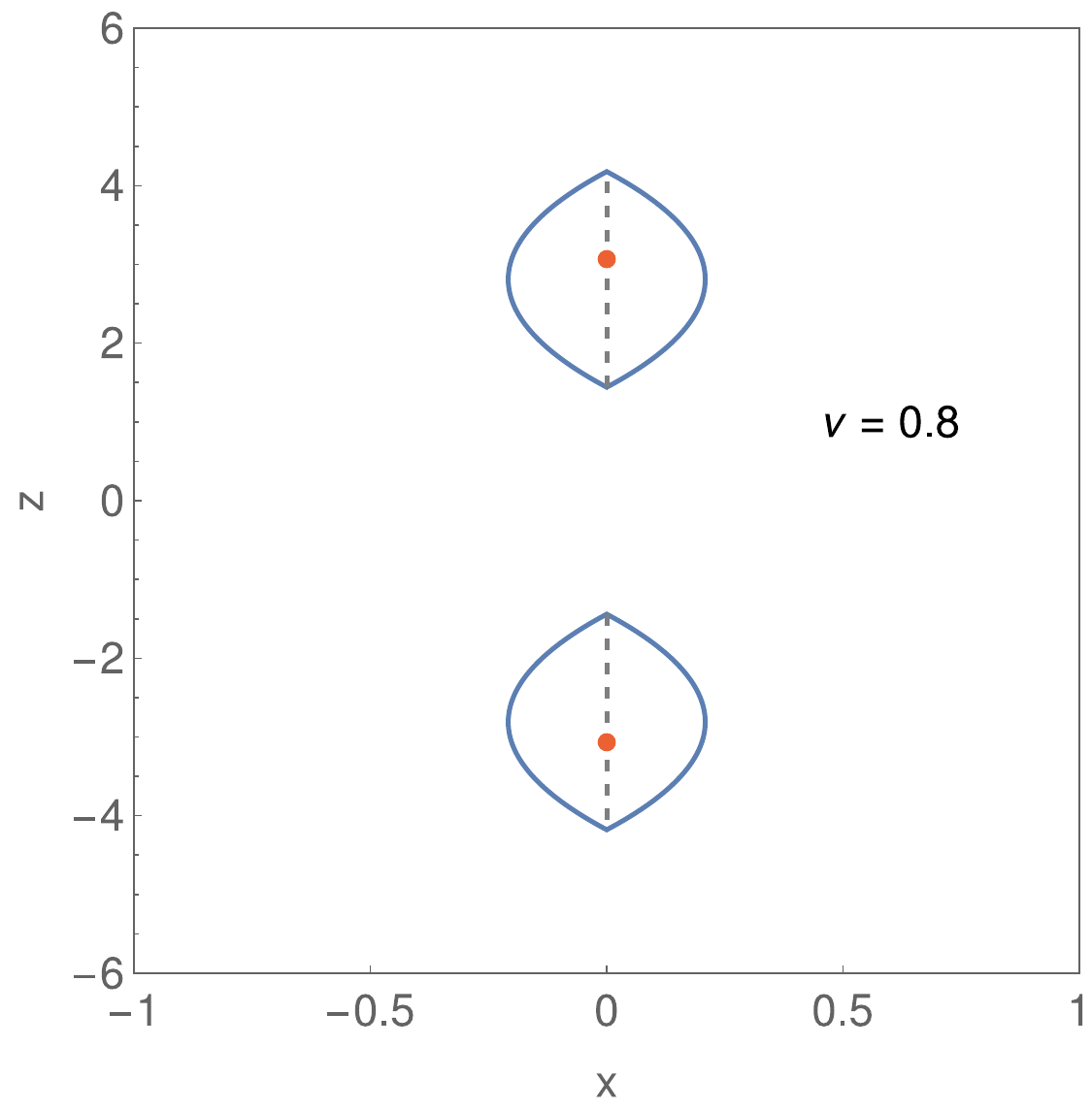}
    \caption{Cut of the spacetime along the $XZ$ plane for $\nu = 0.8$. The orange dots represent the black holes. The blue lines represent the static limit. The dashed grey line represent the horizon rods.}
    \label{fig:horizon8}
\end{minipage}%
\end{figure}

For $\nu = 0$, $\alpha_\pm = k \pm m$ and $d = k m$ are real, and the metric, which is then static, describes a system of two non-rotating black holes each of mass $m$ with a cosmic string of length $2(k- m)$. When $k = m$, the two black holes coalesce back into a single Schwarzschild one.

On the condition that $\alpha_\pm$ and $d$ are real, the system defined by the gravimagnetic dipole spacetime consists of two black holes separated by a distance $2k$ and connected by a spinning cosmic string -- called Misner string -- of length $2\alpha_-$. These correspond to three sections of the $z$ axis; the two black holes horizons for $\alpha_- < z < \alpha_+$ and $-\alpha_+ < z < - \alpha_-$, and the Misner string for $-\alpha_- < z < \alpha_-$. The masses of the black holes $M_{H_\pm}$ and their angular momenta, $ J_{H_\pm}$ are functions of the three parameters $m, k$ and $\nu$ \cite{Clement2021}.

The conditions for $\alpha_\pm$ and $d$ to be real and non zero are given\cite{Clement2021} by
\begin{equation}\label{cond1k}
k>k_+ \mbox{ or } k<k_-, \quad k_\pm(m,\nu) = \sqrt{m^2 + 2 \nu^2} \pm \left\vert\nu \right\vert.
\end{equation}

For the black holes to be at equilibrium with each other, the spinning cosmic string that joins them should be tensionless. The string tension per unit length $(1 - e^{- \gamma_S})/4$ is related to the value of the metric function at its center $e^{2 \gamma_S} \equiv  e^{2 \gamma} (\rho =0, z = 0)$.

This last conditions translates to  
\begin{equation}
\left( \frac{\alpha_+ - \alpha_-}{2 m} \right)^2 = 2 - \left( \frac{2 m}{\alpha_+ + \alpha_-} \right)^2.
\end{equation}
This equation gives, as allowed values for $k$,
\begin{equation}\label{cond2k}
k (m,\nu) = \pm\sqrt{\frac{m^6 + 3 m^4 \nu^2 \pm m^2 \nu \sqrt{4 vm^6 + 9 m^4 \nu^2 + 2 m^2 \nu^4 + \nu^6}}{m^4 - \nu^4}}.
\end{equation}
Assembling eqs.~\eqref{cond1k} and \eqref{cond2k} gives rise to a unique solution for $k$ in the tensionless case:
\begin{equation}
k^T_+(m,\nu) = \sqrt{\frac{m\nu^6 + 3 m^4 \nu^2 + m^2 \nu \sqrt{4 m^6 + 9 m^4 \nu^2 + 2 m^2 \nu^4 + \nu^6}}{m^4 - \nu^4}}.
\end{equation}
Lastly, for $k = m$, it was shown\cite{MankoRodchenko} that the system describes a Kerr black hole of mass $2 m$ and Kerr parameter $\nu = a$, provided that $\nu < 2 m$. In this case, $d = m k$, $\alpha_+$ is real and $\alpha_-$ is imaginary, but $\alpha_-$ will not appear in any of the expressions of the metric, preserving its reality. 

Since the system studied hereafter is the one corresponding to two rotating black holes at equilibrium at a fixed distance, the value $k = k^T_+$ will be used. The system is thus characterized by only two variables, $\nu$ and $m$.

Finally, when the parameters $\nu$, $k$ and $m$, as well as the coordinates $\rho, \phi, z$ and $ct$ are divided by $m$, the metric is fully dimensionless. The symbols will be kept the same for readability's sake, but keep in mind now that every quantity is adimensional; only $m$ will be replaced by $1$.

\subsection{Horizons and ergoregions}

The static limits of the black holes ($g_{tt} = 0$) can be determined and plotted on a XZ slice of the space since it is axisymmetric. The general shape would be of two symmetric pears, or two halves of a peanut shell, that grow further apart as $\nu$ increases. The separation of the black holes $2k$ diverges when $\nu \rightarrow 1$. The horizons are in the shape of rods, for $\rho =0$, $-\alpha_+ < z < - \alpha_-$ and $\alpha_- < z < \alpha_+$. The two horizons touch only in the extreme case $\nu =0$, at the singularity point of coordinates $\rho = 0 = z$, which will not be studied here.

Fig. \ref{fig:horizon_details} shows a cut in the XZ plane of the ergoregions and horizons, as well as the quantities $\alpha_\pm$ (the extremities of the horizons, as well as the intersection between the static limits and the vertical axis) and $k$ (half the black holes separation) for a gravimagnetic dipole spacetime with NUT parameter $\nu = 0.4$.

Figs. \ref{fig:horizon1} and \ref{fig:horizon8} show the horizons and ergoregions for spacetimes with NUT parameters $\nu = 0.1$ and $\nu = 0.8$.

\subsection{The metric functions in the equatorial plane}

Before proceeding let us point out that at $z =0$ the quantities in terms of which the metric components are defined now take the following values:
\begin{equation}
R = R_\pm(\rho, z = 0) = \sqrt{\rho^2 + \alpha_+^2}, \qquad 
r = r_\pm(\rho, z = 0)  = \sqrt{\rho^2 + \alpha_-^2},
\end{equation}
allowing the following definitions 
\begin{equation}\label{eq:funcs_ABG_equatorial_plane}
\begin{aligned}
A_0 & \equiv \frac{A(\rho, z = 0)}{2 \alpha_+ \alpha_-} =   - \Big[(d^2 -1)(R^2 + r^2) + (d^2+ 1)R r \Big], \\
B_0  &  \equiv \frac{B(\rho, z = 0)}{2 \alpha_+ \alpha_-} =   - 2 d \Big[(d-1)R + (d + 1)r \Big], \\
G_0 & \equiv \frac{G(\rho, z = 0)}{2 \alpha_+ \alpha_-} =  G_r + i G_i, \\
G_i & \equiv \Im\left[\frac{G(\rho,z=0)}{2 \alpha_+ \alpha_-}\right] = - k \nu d \left[ (d-1) R + (d+1)r\right],\\
G_r & \equiv \Re\left[\frac{G(\rho,z=0)}{2 \alpha_+ \alpha_-}\right] = -\left[((1 + d) r + (-1 + d) R) (-r + R + d (4 + r + R))\right].
\end{aligned}
\end{equation}
With these redefinitions, it is ensured that $A_0, B_0$, and $G_r$ and $G_i$ are all real functions. They allow for the functions $f$, $\omega$ and $ e^{2 \gamma}$ to be written as
\begin{equation}\label{eq:funcs_fomegae2g_equatorial_plane}
\begin{aligned}
f & \equiv f(\rho,z=0) = \frac{A_0 - B_0}{A_0 + B_0} = -\frac{8 d }{(d-1) r+(d+1) R+4 d}, \\
\omega & = \omega(\rho,z=0) = \frac{- 4 G_i}{A_0 - B_0}=\frac{8 d k \nu}{(1-d) r-(d+1) R+4 d}, \\
e^{2 \gamma} & = e^{2 \gamma} (\rho, z = 0) =  \frac{A_0^2 - B_0^2}{4 d^4 R^2 r^2}=\frac{\left[(d+1) r+(d-1) R\right]^2 \left[\left\{R(d+1)+r(d-1) \right\}^2 - 16 d^2 \right]}{16 d^4 r^2 R^2}.
\end{aligned}
\end{equation}
Every metric function is thus real for any value of $\rho \geq 0$ and $\nu \in \left] 0, 1 \right[$.

\section{The effective potential for geodesics}\label{sec:hamiltonianpotential}

Consider the general case of a free point particle --- be it massive 
or massless --- propagating in a spacetime whose geometry in 
coordinates $(x^0, x^i)$ is characterised by the following metric 
and line element:
\begin{equation}
ds^2 = g_{\mu \nu} d x^\mu d x^\nu.
\end{equation}
From the outset, herein the metric $g_{\mu \nu}$ is taken to be 
stationary (and asymptotically flat). Let us note that $x^0 = ct$ is 
the time coordinate with a dimension of length, and the $x^i$ are spacelike 
(in general, curvilinear) coordinates of that same physical dimension.

As is well  known\cite{GovaertsHam,Polchinski,Beisert}, the particle's 
dynamics is derived from the Hamiltonian first-order action:
\begin{equation}
S[x^\mu,p_\mu ; e] = \int du (\dot{x}^\mu p_\mu - H), \qquad H = 
\frac{1}{2}e(u) (g^{\mu\nu} p_\mu p_\nu + (\mu_0 c)^2),
\end{equation}
with $u \in \mathbb{R}$ being an arbitrary worldline parametrisation 
(the dot standing for a derivative relative to $u$), $\mu_0$ the mass of 
the particle ($\mu_0 > 0$ for a massive particle and $\mu_0 = 0$ for a 
massless one) and $H$ the first-class Hamiltonian.

The conjugate momentum 
$p_0$ is constant due to the stationarity of the metric, and satisfies 
$p_0 = -E / c$, where $E$ is the particle's energy. Additionally, 
when the metric is axisymmetric or spherically symmetric, $p_\phi$ 
is conserved and such that $p_\phi = L$ with $L$ its (orbital) angular 
momentum. The positive definite einbein field $e(u)$ is playing the role 
of the Lagrange multiplier (and pure gauge degree of freedom) for the 
first-class constraint:
\begin{equation}\label{eq:RCgaugecond}
g^{\mu \nu} p_\mu p_\nu + (\mu_0c)^2 =0, \qquad c^2 g^{\mu \nu} p_\mu 
p_\nu + (\mu_0 c^2 )^2 =0.
\end{equation}

The geodesic equations then read,
\begin{equation}\label{eq:RC}
\frac{d x^\mu}{d u} = e g^{\mu \nu} p_\nu, \qquad 
\frac{d p_\mu}{d u} =-\frac{1}{2}e 
\frac{\partial g^{\rho \sigma}}{\partial x^\mu} p_\rho p_\sigma,
\end{equation}
while being subjected as well to the first-class constraint 
eq.~\eqref{eq:RCgaugecond} to ensure invariance under orientation preserving 
worldline diffeomorphisms. Additionally, in the massive case the einbein is determined from the constraint as,
\begin{equation}
e = \pm \frac{1}{\mu_0 c} \sqrt{- g_{\mu \nu} \dot{x}^\mu \dot{x}^\nu},
\end{equation}
the upper sign corresponding to the standard square-root action for a 
relativistic massive particle, such that in a Minkowski spacetime solutions 
of positive energy propagate to the future.

\subsection{The Hamiltonian equations of motion}

Given the gravimagnetic dipole metric components, the general relation between coordinate time and proper time is expressed as,
\begin{equation}
\frac{dt}{d\tau}=\frac{1}{\sqrt{-g_{tt}}} = \frac{1}{\sqrt{f}},\qquad
\frac{d\tau}{dt}=\sqrt{f}.
\end{equation}
Furthermore, in Hamiltonian form, the geodesic equations of motion read as follows, for the spacetime coordinates,
\begin{equation}\label{eqs:dxdu}
c\frac{dt}{du}=e\left(g^{tt}p_t+g^{t\phi}p_\phi\right),\quad
\frac{d\phi}{du}=e\left(g^{t\phi}p_t+g^{\phi\phi}p_\phi\right),\quad
\frac{d\rho}{du}=e\,g^{\rho\rho}p_\rho,\quad
\frac{dz}{du} = e\,g^{zz} p_z,
\end{equation}
and for the conjugate momenta,
\begin{equation}\label{eqs:dpdu}
\frac{dp_t}{du}=0,\qquad
\frac{dp_\phi}{du}=0,\qquad
\frac{dp_\rho}{du}=-\frac{1}{2}e\,\frac{\partial g^{\mu\nu}}{\partial \rho}p_\mu p_\nu,\qquad
\frac{dp_z}{du}=-\frac{1}{2}e\frac{\partial g^{\mu\nu}}{\partial z} p_\mu p_\nu,
\end{equation}
subjected to the following constraint,
\begin{equation}\label{eq:constraint}
g^{tt}p^2_t+2g^{t\phi}p_t p_\phi + g^{\phi\phi} p^2_\phi + g^{\rho\rho} p^2_\rho + g^{zz} p^2_z + (\mu_0 c)^2=0,
\end{equation}
with the conserved conjugate momenta $p_t$ and $p_\phi$,
\begin{equation}
p_t=-\frac{E}{c},\qquad p_\phi=L,
\end{equation}
where $E$ is the particle's relativistic energy and $L$ its angular momentum (component along the axial symmetry axis), both taking constant and real values.

\subsection{The effective potential}

In the equatorial plane, $p_z = 0 = z$\footnote{See Appendix \ref{gravidip_condition_z} for the verification of $d p_z/du = 0$, which is a non-trivial property.} and the Hamiltonian of a particle can be expressed, be it massive ($\mu_0 > 0$) or massless ($\mu_0 = 0$), as
\begin{equation}\label{eq:hamiltonian}
\mathcal{H}  = \frac{e}{2} \frac{1}{c^2} \left(g^{tt} E^2 - 2g^{t\phi} E (cL) + g^{\phi \phi} (cL)^2 + g^{\rho \rho} p_{\rho}^2+ (\mu_0 c^2)^2\right).
\end{equation}
This expression will now be used to define an effective potential to study circular orbits. From the definition of the conjugate momenta, $p_\rho$ is expressed in terms of $d \rho / d \phi$ as 
\begin{equation}
p_\rho = g_{\rho \rho} \frac{d \rho}{d \phi} \frac{d \phi}{du}  = g_{\rho \rho} \left( g^{t\phi} p_t + g^{\phi \phi} p_\phi \right) \frac{d \rho}{d \phi}.
\end{equation}

The constraint in eq.~\eqref{eq:constraint} can then be reexpressed as 
\begin{equation}
\left( \frac{d \rho}{d \phi} \right)^2 + V(\rho, b) = 0,
\end{equation}
where
\begin{equation}\label{eq:potential}
V(\rho, b) = \frac{\rho^2}{e^{2 \gamma} } \left[ 1 - \frac{\rho^2}{f^2} \frac{1}{(b-\omega)^2} \left( 1 - f \left(\frac{\mu_0 c^2}{E} \right)^2 \right)\right],
\end{equation}
and $b = cL/E$ is the impact parameter\cite{Misner}.

Circular orbits correspond to the double condition that $V(\rho) = 0$ and $V'(\rho) = 0$ (the $'$ here denoting the derivative taken with respect to $\rho$). These equations are nontrivial but the first one can be used to define a critical impact parameter $b_\pm$, which allows to simplify the expression of the second one:
\begin{equation}
b_\pm = \omega \pm \frac{\rho}{f} \sqrt{1 - f \left(\frac{\mu_0 c^2}{E} \right)^2}, \qquad V(\rho, b = b_\pm) = 0.
\end{equation}
The $\pm$ sign factor corresponds to the direction of rotation of the particle on its orbit, prograde or retrograde respectively. Again, this expression is valid for massive ($\mu_0>0$) or massless ($\mu_0 = 0$) particles.

The derivative of the potential with respect to $\rho$ can be written as the sum of two terms:
\begin{equation}
V'(\rho, b) \Big\vert_{b = b_\pm} = \left(\frac{\rho^2}{e^{2 \gamma}}\right)' \underbrace{\Bigg[ ... \Bigg]}_{=0 \mbox{ for } b = b_\pm} + \frac{\rho^2}{e^{2 \gamma}}\underbrace{\Bigg[ ... \Bigg]'}_{= g(\rho)}.
\end{equation}
The first is automatically zero by choice of $b = b_\pm$; the second term needs a bit more work. Since $e^{2 \gamma} \neq 0$ for any value of $(\nu, \rho)$ such that $(\nu>0, \rho>0)$, the expression of interest will be only the large parenthesis that will be called $g(\rho)$ in the following.

\subsection{Massive particules}
Writing out the complete expression of $g(\rho)$ gives:
\begin{equation}\label{eq:rhomassive}
\begin{aligned}
g(\rho) & = \frac{\rho \left[(b_\pm-\omega) (\rho f' (2 E^2-c^4 \mu_0^2 f)+2 f (c^4 \mu_0^2 f-E^2))+2 \rho f \omega' (c^4 \mu_0^2 f-E^2)\right]}{E^2 f^3 (b_\pm-\omega)^3} \\
& = \frac{2 f^2 (c^4 \mu_0^2 \mp \beta E^2 \omega')-f (c^4 \mu_0^2 \rho f'+2 E^2)+2 E^2 \rho f'}{\beta^2 E^2 \rho f}, \\
&  \mbox{ with } \beta =  \sqrt{1 - f \left(\frac{\mu_0 c^2}{E} \right)^2}, \, b_\pm = \omega \pm \frac{\rho}{f} \beta.\\
\end{aligned}
\end{equation}

The radii $\rho_\pm$ of the circular orbits are then given by the solutions of $g(\rho) = 0$, the sign factor corresponding to the one of $b_\pm$ and thus either a prograde or retrograde orbit. According to the value of  the the second derivative in $\rho$ of the potential, $V''$, this orbit will be either stable ($V''<0$) or unstable ($V''>0$). Depending on the values of the NUT parameter $\nu$ and of the energy $E$, eq.~\eqref{eq:rhomassive} will have between four and zero possible solutions.

\begin{figure}
\centering
\includegraphics[width=0.7\textwidth]{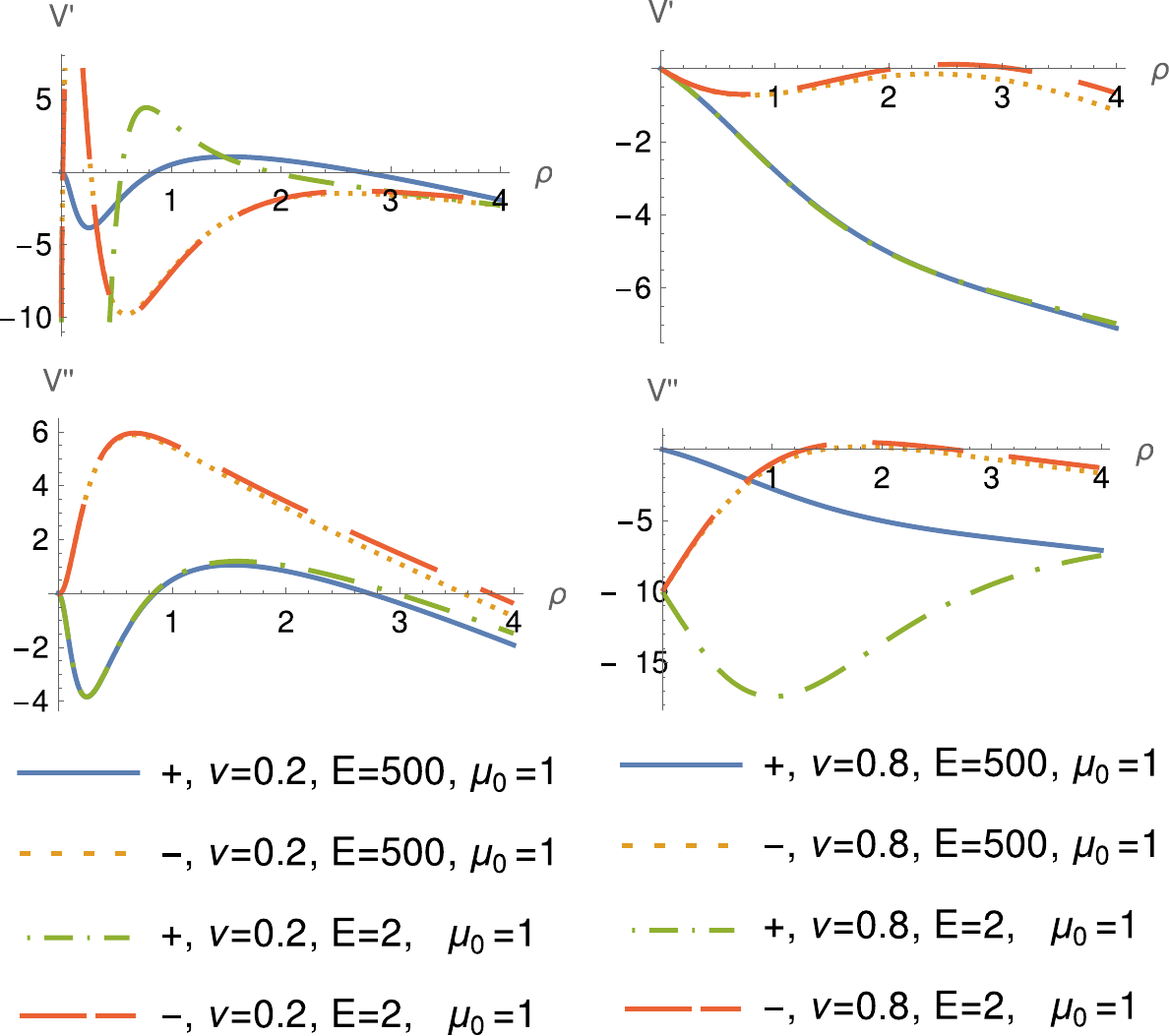}
\caption{\label{figpotentials_massive} The left column has a NUT parameter $\nu = 0.2$ and the right column $\nu = 0.8$. The first (resp. second) row presents the first (resp. second) derivative $V'$ (resp. $V''$) of the potential as functions of $\rho$ and evaluated at $b = b_\pm$. Two values of the energy are represented ($E = 2$ and $E = 500$) for the prograde and retrograde directions. The zeroes of the first derivative of the potential indicate the radii at which circular orbits are possible, and the value of the second derivative at these radii indicate if the corresponding orbit is stable or unstable. For example, for $\nu = 0.2$, with an energy of $E = 2$ (green dashdotted line on the left column), there is a stable prograde orbit at $\rho \approx 0.5$.}
\end{figure}

In Fig. \ref{figpotentials_massive}, the left column has a NUT parameter $\nu = 0.2$ and the right column $\nu = 0.8$. The first (resp. second) row presents the first (resp. second) derivative of the potential, as function of $\rho$ and evaluated at $b = b_\pm$. For each graph, two values of the energy are represented ($E = 2$ and $E = 500$) for the prograde and retrograde directions. The zeroes of the first derivative of the potential indicate the radii at which circular orbits are possible, and the value of the second derivative at these radii indicate if the corresponding orbit is stable or unstable. For example, for $\nu = 0.2$, with an energy of $E = 2$, (green dashdotted line on the left column) there is a stable prograde orbit at $\rho \approx 0.5$.

\begin{figure}
\centering
\includegraphics[width=\textwidth]{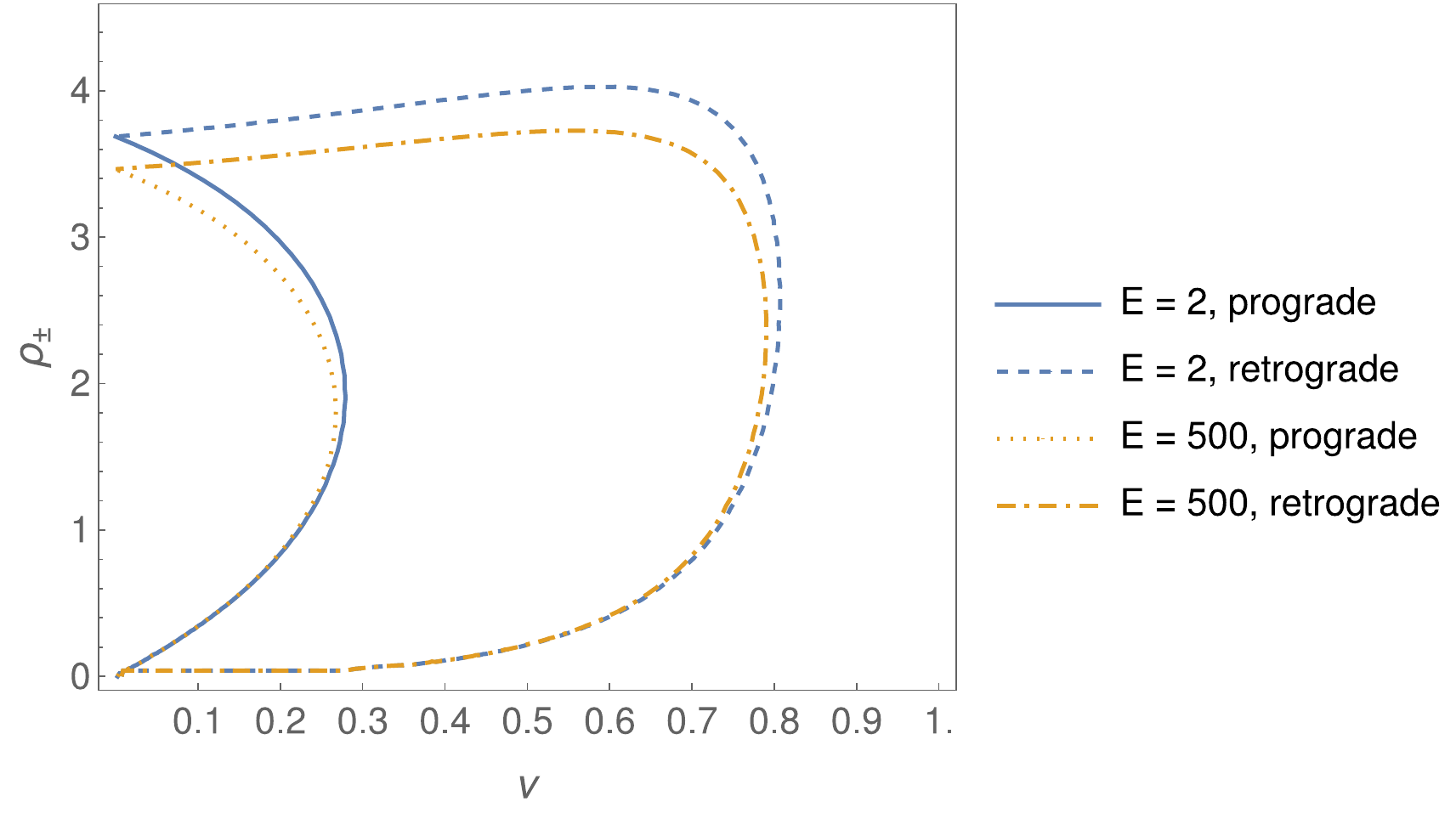}
\caption{\label{fig:circular_radii} For different values of the energy $E$, as functions of the parameter $\nu$, the radii of the possible circular orbits $\rho_\pm$ for massive particles. Depending on the value of the NUT parameter $\nu$ and the energy $E$, there is from $0$ to $4$ possible orbits.}
\end{figure}

Eq.~\eqref{eq:rhomassive} can also be used to construct a bifurcation diagram, providing a visualisation of how the number of solutions (\textit{i.e.} the number of possible circular orbits) varies with the value of $\nu$. This is done in Fig. \ref{fig:circular_radii} for two different values of the energy. As the value of $\nu$ grows, the number of circular orbits decreases. The prograde (resp. retrograde) orbits coalesce into one prograde (resp. retrograde) orbit, then disappear.

\subsection{Massless particles}

In the case of massless particles, $b_\pm = \omega \pm \frac{\rho}{f}$ and $\beta = 1$, and eq. \eqref{eq:rhomassive} can be simplified to
\begin{equation}\label{eq:rhomassless}
g(\rho) = -\frac{2 ( \pm f^2 \omega'+f -f'\rho )}{f \rho}.
\end{equation}

\begin{figure}
\centering
\includegraphics[width = 0.5\textwidth]{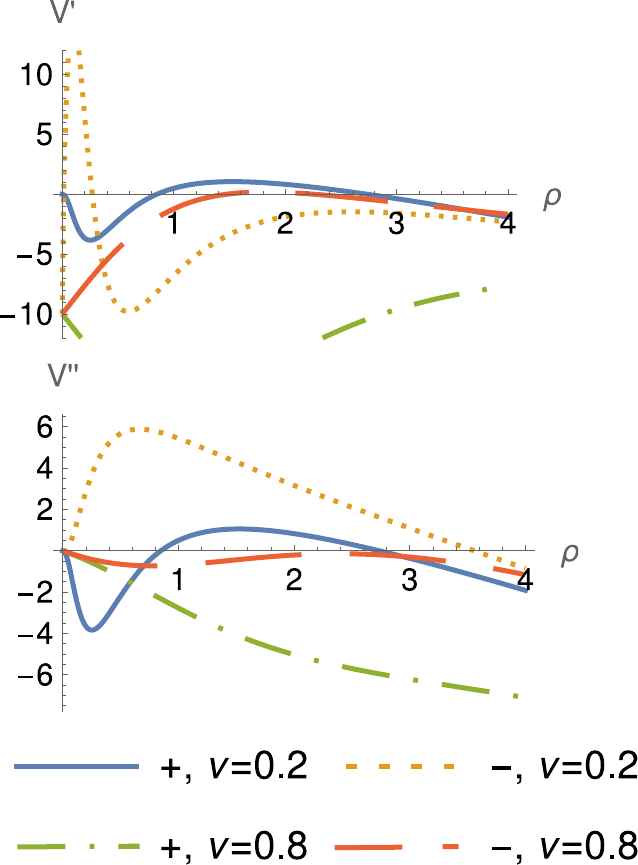}
\caption{\label{figpotentials_massless} The top (resp. bottom) graph represents the first (resp. second) derivative of the potential as a function of $\rho$ for massless particles, for two different rotation directions (prograde and retrograde) and two different values of $\nu$. The number of possible circular orbits (from $0$ to $4$) depends on this parameter.}
\end{figure}

The radii $\rho_\pm$ of the photon orbits are then given by the solutions of $g(\rho) = 0$, the sign factor corresponding to the one of $b_\pm$ and thus either a prograde or retrograde orbit. According to the value of  the the second derivative in $\rho$ of the potential, $V''$, this orbit will be either stable ($V''<0$) or unstable ($V''>0$). Depending on the values of the NUT parameter $\nu$ and of the energy $E$, eq.~\eqref{eq:rhomassless} will have between four and zero possible solutions.

In Fig. \ref{figpotentials_massless}, the top (resp. bottom) graph presents the first (resp. second) derivative of the potential for a massless particle as function of $\rho$ and evaluated at $b = b_\pm$. For each graph, two values of the NUT parameter ($\nu = 0.2, \nu = 0.8$) are represented for the prograde and retrograde directions. The zeroes of the first derivative of the potential indicate the radii of the photon orbits, and the value of the second derivative at these radii indicate if the corresponding orbit is stable or unstable. For example, for $\nu = 0.2$ (orange dotted line) there is a unstable retrograde orbit at $\rho \approx 0.3$.

\begin{figure}
\centering
\includegraphics[width=0.8\textwidth]{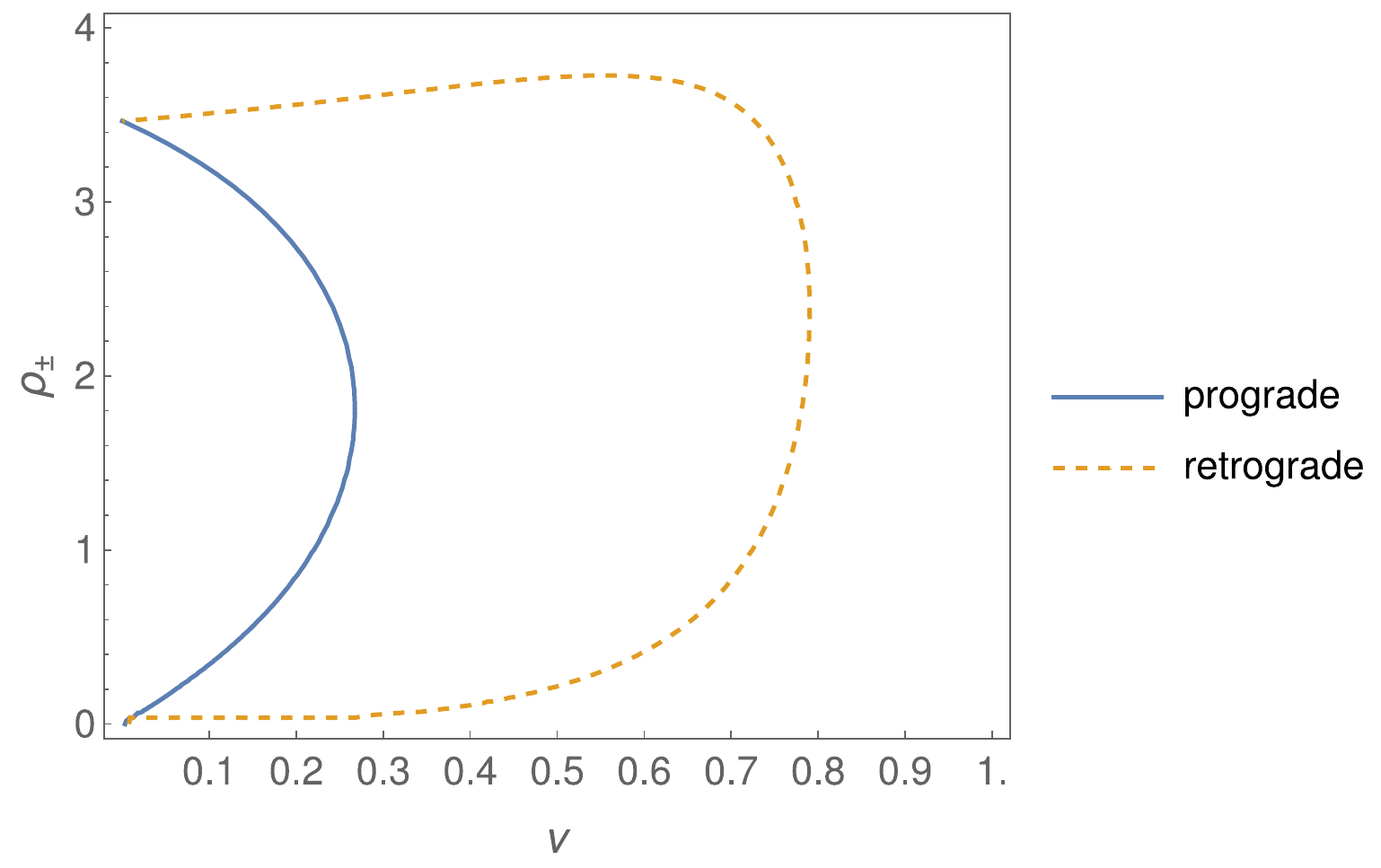}
\caption{\label{fig:circular_radii_massless} As functions of the parameter $\nu$ for massless particles, the radii of the possible circular orbits $\rho_\pm$ are represented. The plain line corresponds to prograde orbits, the dashed one to retrograde orbits. Depending on the value of $\nu$, there will be between zero and four possible circular orbits.}
\end{figure}

Like in the massive case, eq.~\eqref{eq:rhomassless} can be used to construct the same kind of bifurcation diagram, providing a visualisation of how the number of solutions varies with the value of $\nu$. This is done in Fig. \ref{fig:circular_radii_massless}. As the value of $\nu$ grows, the number of photon orbits decreases. The prograde (resp. retrograde) orbits coalesce into one prograde (resp. retrograde) orbit, then disappear. The coordinates of the saddle-node bifurcation can be numerically computed: for the prograde orbit, $\nu = 0.27, \rho = 1.78$, and for the retrograde orbit $\nu = 0.79, \rho = 2.30$.

\FloatBarrier
\section{Proper distance and velocity}\label{sec:distancevelocity}

The radar distance, or spatial distance\cite{LandauLifshitz3} is the distance determined by the following procedure: an observer at point $B$ sends a light signal to  point $A$ and receives its reflection back after some short interval of time. The radar distance between $A$ and $B$ is defined as half the total travel time (measured in the observer's proper time) multiplied by the speed of light.

This is expressed as
\begin{equation}
d\ell^2 = \left(g_{ij} - \frac{g_{0i}g_{0j}}{g_{00}}\right) dx^i dx^j.
\end{equation}

This expression differs from the purely spatial 
distance by the term $\displaystyle{\frac{-g_{0i} g_{0j}}{g_{00}}} dx^i dx^j$,
which takes into account the spatial deformation due to the advancement of 
time (note that it is also, of course, of the right dimension). It can be 
rewritten in the form:
\begin{equation}\label{def_dist}
d\ell^2 = \gamma_{ij} dx^i dx^j, \qquad \mbox{with} \qquad 
\gamma_{ij} = g_{ij} - \frac{g_{0i} g_{0j}}{g_{00}}.
\end{equation}

The tensor $\gamma_{ij}$ is the reciprocal of the 
contravariant spatial tensor $g^{ij}$. Indeed, from the relation 
$g^{\alpha \gamma} g_{\gamma \beta} = \delta^\alpha_\beta$, it follows that
\begin{equation}
g^{ij}g_{jk} + g^{i0}g_{0k} = \delta^i_k, \qquad g^{i0} = 
-\frac{g^{ij} g_{j0}}{g_{00}},
\end{equation}
and by inserting $g^{i0}$ in the first of these two relations, one finds,
\begin{equation}
g^{ij} \gamma_{jk} = \delta^i_k.
\end{equation}

Since in the general case $g_{\mu \nu}$ depends on $x^0$, 
it is meaningless to integrate $d\ell$ given a generic curved spacetime. 
The integral would be ill-defined since it would depend on the worldline, 
{\sl i.e.}, the chosen path between the two points. Thus, generally 
speaking, in the context of generic curved spacetimes the concept 
of spatial distance remains valid at best only for infinitesimally 
small distances.

However, in the case of a stationary metric, $g_{\mu \nu}$ does not 
depend on $x^0$ and as such, the integral $\int d\ell$ is well-defined 
and can be used to determine the finite spatial distance between two 
simultaneous events. 

Let us also note that in the axisymmetric spacetime under consideration here, the only cross term is $g_{t\phi}$ and there is no $g_{t \rho}$ crossterm. As such, the spatial radial distance corresponds to the proper distance $\bar{\rho}$ between $\rho$ and $\rho_0$ defined as
\begin{equation}
\bar{\rho} = \int^{\rho}_{\rho_0} \sqrt{g_{\rho \rho}} \, d\rho.
\end{equation}

Now that the concept of spatial distance is properly defined, the proper 
velocity (also known as celerity) can be defined as follows relative 
to the proper time along the worldline:
\begin{equation}
v_\tau^2 \equiv \gamma_{ij} \frac{dx^i}{d\tau} \frac{dx^j}{d\tau},
\end{equation}
as well as the velocity relative to the coordinate time, or coordinate 
velocity:
\begin{equation}\label{eq:velocity_def}
v_t^2 \equiv \gamma_{ij} \frac{dx^i}{dt} \frac{dx^j}{dt} = 
\left( \frac{d\tau}{dt} \right)^2 v_\tau^2, \qquad v_\tau^2 = 
\left(\frac{dt}{d\tau}\right)^2 v_t^2.
\end{equation}

The coordinate velocity, denoted as $v_t>0$ in the following, 
represents the particle's velocity as recorded by an asymptotic 
observer\footnote{In view of the
physical context, it is implicitly assumed herein that the stationary 
metric is asymptotically flat.}. The proper velocity, $v_\tau>0$, 
represents the ratio between the observer-measured displacement and the proper time elapsed on the clocks of the particle. In the following, focus is being put on the coordinate 
velocity as must be measured by an asymptotic observer, hence its 
physical relevance to this article.

\subsection{Proper distance in the gravimagnetic dipole spacetime}
On the equatorial plane $z = 0$, the $g_{\rho \rho}$ factor can be written quite nicely:
\begin{equation}
\frac{e^{2 \gamma}}{f} = \frac{\lvert A + B \rvert ^2}{64 \alpha_+^2 \alpha_-^2 d^4 R^2 r^2}.
\end{equation}
When divided by $\alpha_+\alpha_-$, at $z = 0$, $A + B$ is real and negative. As such,
\begin{equation}
\begin{aligned}
\sqrt{g_{\rho \rho}} & =  - \frac{A + B}{\alpha_+ \alpha_-} \frac{1}{8 d^2 R r} \\
& =  \frac{1}{4}\left( 1-\frac{1}{d^2}\right) \left(\frac{r}{R}+\frac{R}{r}\right)+\frac{1}{2 d^2}+\frac{1}{d} \left(\frac{1}{R}-\frac{1}{r}\right)+ \left(\frac{1}{R}+\frac{1}{r}\right)+\frac{1}{2}.
\end{aligned}
\end{equation}

This expression can be integrated in the following manner:
\begin{equation}
\begin{aligned}
& \int \sqrt{g_{\rho\rho}} \, d\rho = \\
& \frac{- i (d^2 - 1)}{4 d \alpha_+} \left\{ \alpha_+^2 E \left( i \arcsinh \left(\frac{\rho}{\alpha_-}\right), \frac{\alpha_-^2}{\alpha_+^2}\right) + (\alpha_-^2 - \alpha_+^2) F\left( i \arcsinh\left( \frac{\rho}{\alpha_-}\right), \frac{\alpha_-^2}{\alpha_+^2}\right)\right\} \\
& + \frac{1}{d^2} \left\{ \frac{\rho}{2} \left( d^2 + 1 \right) + d  \left[ (d-1) \log(\rho + r) + (d + 1) \log(\rho + R)\right] \right\},
\end{aligned}
\end{equation}
where $F(\phi,m)$ and $E(\phi, m)$ are the elliptic integrals of the first and second kind defined as: 
\begin{equation}
\begin{aligned}
E(\phi, m) & = \int^\phi_0(1 - m^2 \sin^2\theta)^{1/2} \, d\theta, \\
F(\phi, m) &  = \int^\phi_0 (1 - m^2 \sin^2\theta)^{-1/2} \, d\theta.
\end{aligned}
\end{equation}

\begin{figure}
\centering
\includegraphics[width = 0.75\textwidth]{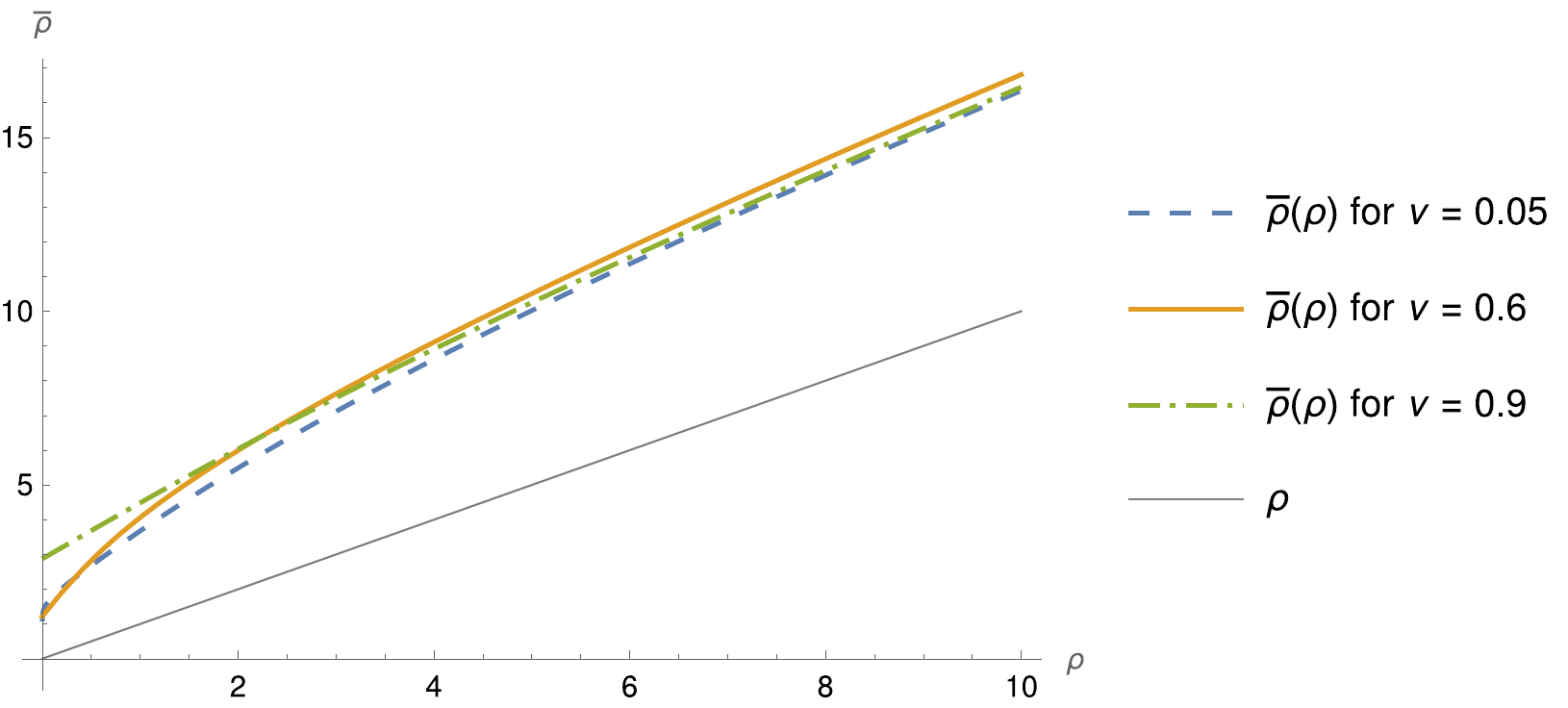}
\caption{\label{fig:properdistance} The proper radius for three values of the NUT parameter is plotted as a function of the coordinate radius. As a reference, the thin grey line is also the coordinate radius as a function of the coordinate radius, \textit{i.e.} the identity.}
\end{figure}

The proper radius between $\rho =0$ and the position of a particle at coordinate $\rho$ is then defined as 
\begin{equation}
\bar{\rho} = \int^\rho_0 \sqrt{g_{\rho\rho}} \, d\rho.
\end{equation}
In Fig. \ref{fig:properdistance}, the proper radii for three different values of $\nu$ are plotted as function of $\rho$. It is not immediately obvious on the graph, but 
\begin{equation}
\lim_{\rho \rightarrow \infty} \bar{\rho}(\rho) = \rho.
\end{equation}
Let us insist that $\bar{\rho}(\rho)$ depends not only on $\rho$, but also on the value of the NUT parameter $\nu$.

\subsection{Velocities for circular orbits in the gravimagnetic dipole spacetime}\label{sec:velocities}

For a circular rotation curve in the equatorial plane, we require $d \rho /du = 0$ and $z = 0,$ $dz/du = 0$, meaning that the velocity can be written from eqs. \eqref{eqs:dxdu} and \eqref{eq:velocity_def} as:
\begin{equation} \label{eq:velocity}
\begin{aligned}
v_t^2 & = \left( g_{\phi \phi} - \frac{(g_{t\phi})^2}{g_{tt}} \right) \left( \frac{d \phi}{dt}\right)^2\\
& =  \left( g_{\phi \phi} - \frac{(g_{t\phi})^2}{g_{tt}} \right) \frac{g^{\phi \phi} b - g^{t \phi}}{g^{t \phi} b - g^{tt}} \\
\Rightarrow \frac{v_t^2}{c^2} & = \frac{\rho^2 f^3 (b-\omega)^2}{(\rho^2 + f^2 \omega (b - \omega))^2} , \mbox{ where } b = b_\pm = \omega \pm \frac{\rho}{f} \sqrt{1 - f \left( \frac{\mu_0 c^2}{E} \right)^2}.
\end{aligned}
\end{equation}

The expressions of the potentials give a nice representation of where the circular orbits might be, but are not very practical to compute the velocities since one has to solve
eq.~\eqref{eq:rhomassive} (numerically) to find the energy needed for a circular orbit for each value of the radius. As such, hereafter another approach is being used to determine the energy corresponding to a particular value of $\rho$.

\subsection{The massive case}

Replacing the value of $b_\pm$ obtained from eq.~\eqref{eq:rhomassive} in eq.~\eqref{eq:velocity}, the velocity of a circular orbit as a function of the radius is given by
\begin{equation} 
\frac{v_t^2}{c^2}  = \frac{\beta^2 \rho^2 f}{(\rho \pm \beta f \omega )^2}, \quad \mbox{ where } \beta =  \sqrt{1 - f \left(\frac{\mu_0 c^2}{E} \right)^2}.
\end{equation}

As was mentioned earlier, the value of the energy $E$ in this last equation can be found for each value of $\rho$ by solving eq. \eqref{eq:rhomassive}, but can also be expressed in a simpler manner.

One of the conditions for a circular orbit is $dp_\rho/du = 0$ and  eq. \eqref{eqs:dpdu} can be used to rewrite it as
\begin{equation}
 \partial_\rho g^{tt}  - 2 \partial_\rho g^{t \phi} \frac{cL}{E} + \partial_\rho g^{\phi\phi} \left(\frac{cL}{E}\right)^2 = 0.
\end{equation}
In conjunction with the constraint \eqref{eq:constraint}, this gives the expressions for the energy $E$ and the angular momentum $L$ for a circular orbit of radius $\rho$:
\begin{equation}\label{eq:ELvals}
\left\{
\begin{aligned}
E^2 & = \frac{-(\mu_0 c^2)^2 }{g^{tt}  - 2 g^{t \phi} b + g^{\phi \phi} b^2},  \\
(cL)^2 & = \frac{-(\mu_0 c)^2 b^2}{g^{tt} - 2 g^{t \phi} b  + g^{\phi \phi} b^2},
\end{aligned}\right.
\end{equation}
with 
\begin{equation}
\frac{cL}{E} = b = \frac{\partial_\rho g^{t\phi} \pm \sqrt{\left( \partial_\rho g^{t\phi}\right)^2 - \partial_\rho g^{\phi\phi} \partial_\rho g^{tt}}}{\partial_\rho g^{\phi\phi}}.
\end{equation}

\begin{figure}
\centering
\includegraphics[width = \textwidth]{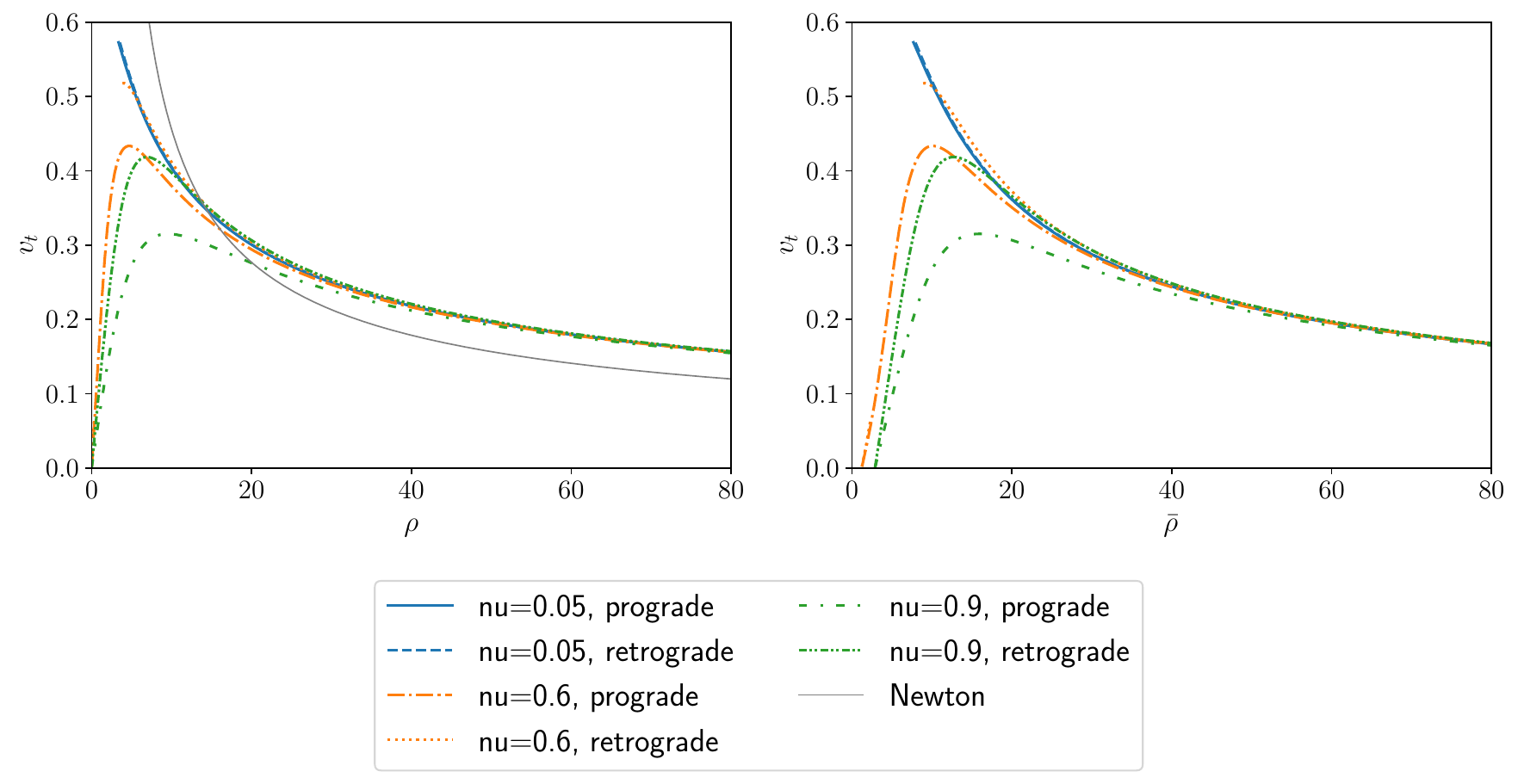}
\caption{\label{fig:velocity_massive} The velocity of a massive particle on a circular orbit is plotted as a function of, on the left-most (resp. right-most) plot, the radius $\rho$ (resp. the proper radius $\bar{\rho}$), for different values of the NUT parameter $\nu$. For the retrograde orbits, the opposite of the velocity is plotted for clarity's sake.  The discontinuities (notably for the prograde and retrograde direction for $\nu = 0.05$ and for the retrograde direction for $\nu = 0.6$) appear due to the presence of photon orbits at these radii; the energy of a massive particle should be infinite. This phenomenon does not occur for $\nu = 0.9$ since there are no photon orbits for this value of the NUT parameter. In between the photon orbits, the energy of a massive particle on a circular orbit should be negative, as shown on Fig. \ref{fig:energy_massive}. The Newtonian velocity curve, $\rho^{-1/2}$, is also displayed on the left-most plot for comparison. }
\end{figure}

\begin{figure}
\centering
\includegraphics[width = \textwidth]{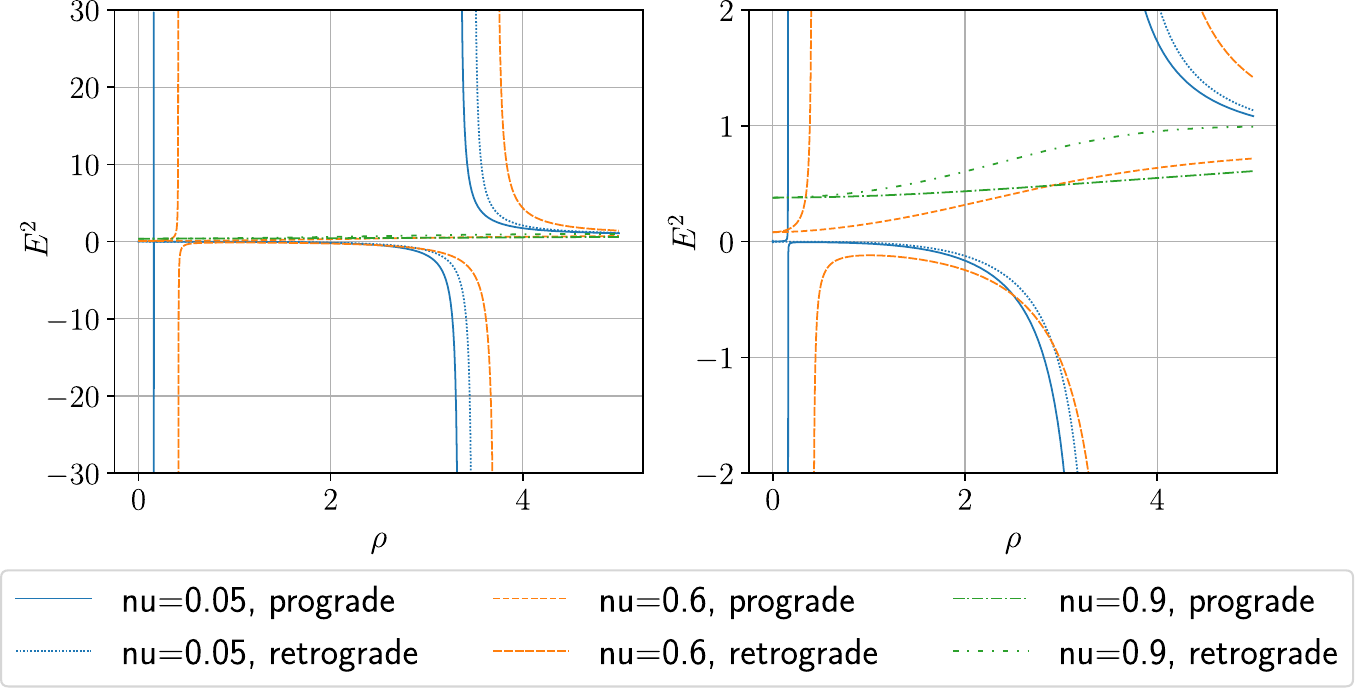}
\caption{\label{fig:energy_massive} The squared value of the energy (eq. \eqref{eq:ELvals}, with $\mu_0 = c = 1$) is plotted as a function of the radius $\rho$ for (prograde or retrograde) circular orbits and for different values of the NUT parameter. The asymptotes towards positive infinity mark the radii at which photon orbits are permitted. In between, the energy is imaginary and there are no circular orbits.}
\end{figure}

Fig. \ref{fig:velocity_massive} shows the velocity of a massive particle on a circular orbit as a function of, on the leftmost (resp. rightmost) plot, the radius $\rho$ (resp. the proper radius $\bar{\rho}$), for different values of the NUT parameter $\nu$. For the retrograde orbits, the opposite of the velocity is plotted for clarity's sake. The discontinuities (notably for the prograde and retrograde direction for $\nu = 0.05$ and for the retrograde direction for $\nu = 0.6$) appear due to the presence of photon orbits at these radii; the energy of a massive particle should be infinite. This phenomenon does not occur for $\nu = 0.9$ since there are no photon orbits for this value of the NUT parameter. In between the photon orbits, the energy of a massive particle on a circular orbit is imaginary (eq. \eqref{eq:ELvals}), as shown on Fig. \ref{fig:energy_massive}. The Newtonian velocity curve, $\rho^{-1/2}$, is also displayed on the left-most plot for comparison. 

\begin{figure}
\centering
\includegraphics[width=\textwidth]{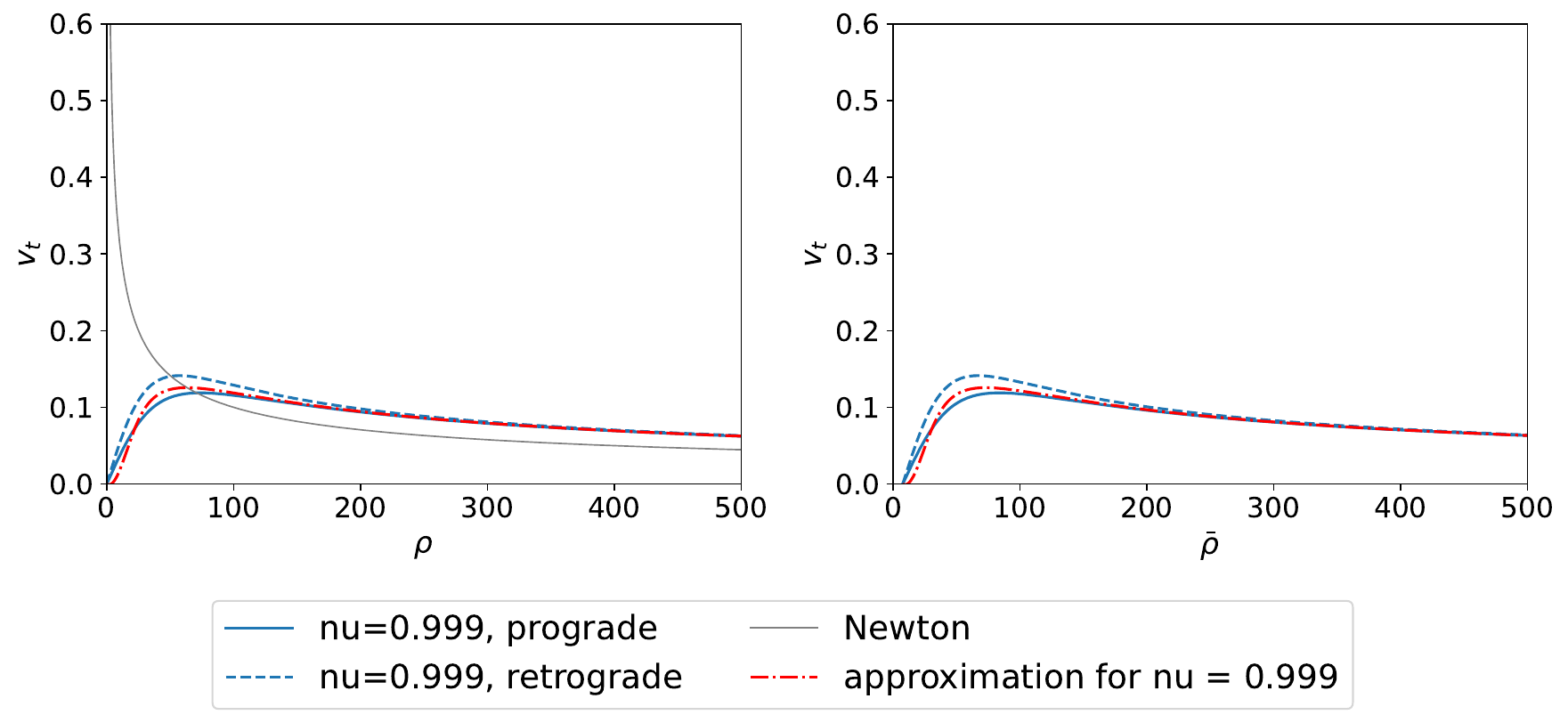}
\caption{\label{fig:comp_approx} For a value of the NUT parameter of $\nu=0.999$, the blue plain (resp. dashed) line depicts the exact velocity of a massive particle on a prograde (resp. retrograde) circular orbit. The red dashdotted line depicts the gravito-electromagnetic approximation in eq.~(79) of Ref. \cite{Govaerts2023}.}
\end{figure}

In Fig. \ref{fig:comp_approx} the velocity rotation curve, computed at $\nu = 0.999$, is compared with the approximate expression given in eq.~(79) of \cite{Govaerts2023}. In this regime ($\nu \approx 1$, tensionless Misner string), where the gravito-electromagnetic approximation is expected to hold, we find close agreement between the two results. (For $\nu = 0.999$ the following values are found: $k = 44.716$, $\alpha_+ = 46.107, \alpha_- = 43.279$.)

\subsection{The massless case}

In the massless case, the value of the radius of a photon orbit for some value of the NUT parameter is given by the solution of eq.~\eqref{eq:rhomassless}. Replacing  $b_\pm = \omega \pm \frac{\rho}{f}$ in eq.~\eqref{eq:velocity}, the velocity of the photons on their circular orbit is given by
\begin{equation}
\frac{v_{t, \pm}^2}{c^2} = \frac{\rho^2 f}{(\rho \pm f \omega)^2},
\end{equation}
where all functions are evaluated at $\rho = \rho_\pm$. 

\begin{figure}
\centering
\includegraphics[width = 0.8\textwidth]{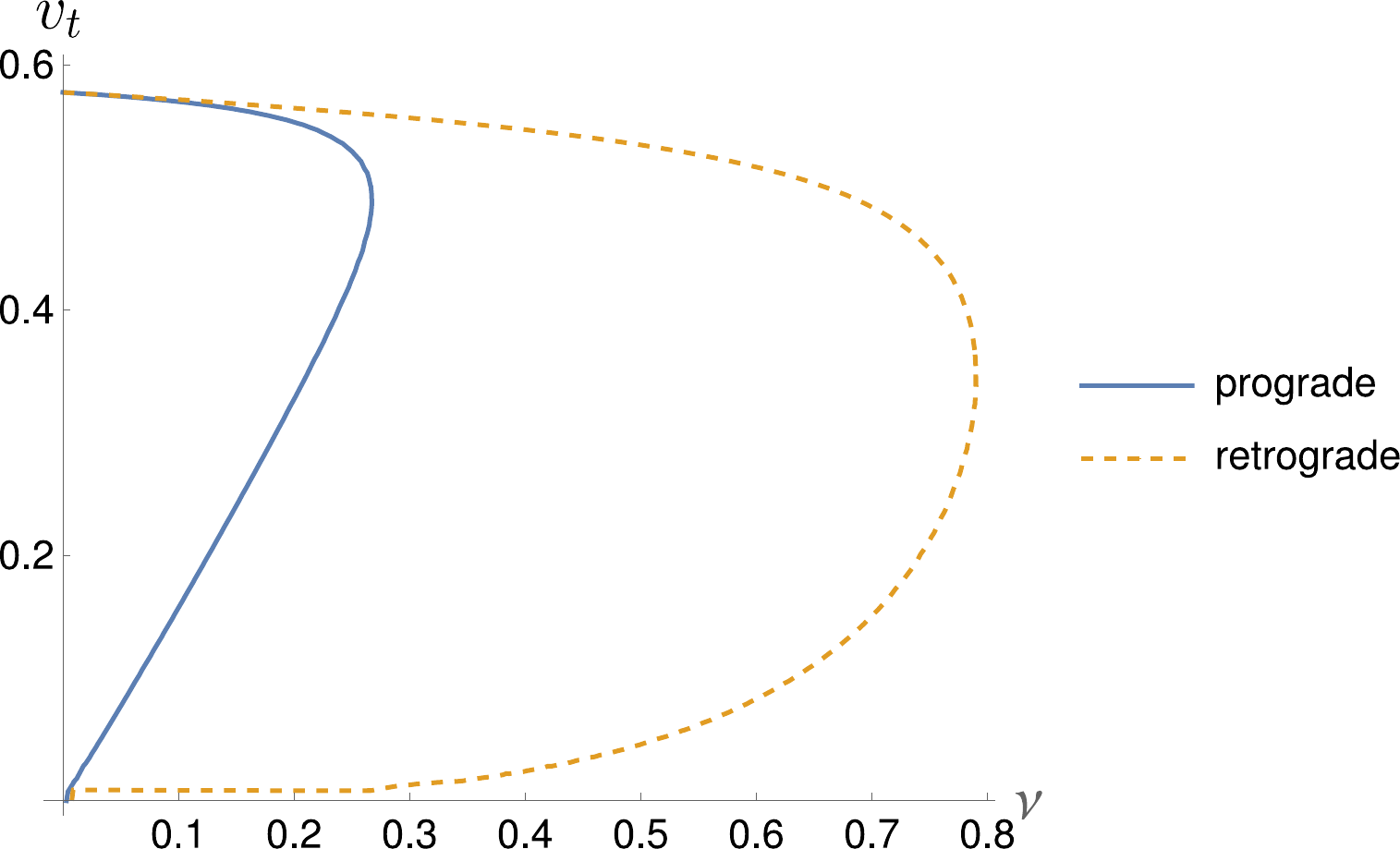}
\caption{\label{fig:velocity_massless} As functions of the NUT parameter $\nu$, the velocities $v_t$ for massless particles on circular orbits. The plain line corresponds to the prograde velocities, the dashed line to the opposite of the retrograde velocity.}
\end{figure}

This velocity is plotted as a function of the NUT parameter $\nu$ on Fig. \ref{fig:velocity_massless}. As should be expected, the figure has again the shape of a bifurcation diagram. When the prograde (resp. retrograde) orbits coalesce together, the corresponding velocities also coalesce together.

\section{Conclusion}

This paper has presented a general method to compute the velocity of a massive, or massless, particle on a circular orbit in the equatorial plane of an axisymmetric stationary and asymptotically flat metric. The approach has then been applied specifically to the gravimagnetic dipole metric, allowing for an analysis of circular rotation curves for both massive and massless particles.

The next natural step would be to investigate the behaviour of particles not confined strictly to the equatorial plane. For example, if particles start in the neighbourhood of the equatorial plane but not in it, they should manifest an oscillatory behaviour around it. This phenomenon could be explored through perturbative methods.

Furthermore, the model could be enriched by adding matter distributions, superposed on the gravimagnetic dipole metric. Such modifications would allow for a more realistic representation of astrophysical scenarios and would very likely influence the shape of the rotation curves.

\appendix

\section{Condition in $z$ is verified $\forall p_t, p_\phi$ in $z = 0$}\label{gravidip_condition_z}

The following equation should be trivially satisfied in $z = 0$ for any $p_t$ and $p_\phi$:
\begin{equation}\label{cond_z_app}
\begin{aligned}
\frac{d p_z}{du} & = - \frac{1}{2} e \frac{\partial g^{\mu \nu}}{\partial z} p_\mu p_\nu \\
0 & = \partial_z g^{tt} p_t^2 + 2 \partial_z g^{t \phi} p_t p_\phi + \partial_z g^{\phi \phi} ,
\end{aligned}
\end{equation}
and here are the derivatives in $z$ of the metric:
 \begin{equation} \label{derivatives}
\left\{\begin{aligned}
\partial_z g^{tt} & = f^{-2} \partial_z f + \frac{1}{\rho^2} \left(\omega^2 \partial_z f  + 2 f \omega \partial_z \omega \right), \\
\partial_z g^{t\phi} & =  \frac{1}{\rho^2} \left(\omega^2 \partial_z f + 2 f \omega \partial_z \omega\right), \\
\partial_z g^{\phi\phi} & = \frac{1}{\rho^2} \partial_z f.
\end{aligned}\right.
\end{equation}

For eq. \eqref{cond_z_app} to be verified $\forall \rho>0, z = 0$, using  \eqref{derivatives}, we need only to verify that for $\rho >0, z = 0$, $\partial_z f = 0$ and $\partial_z \omega = 0$.

Let us start by noticing the following relations when $z = 0$:
\begin{equation}
\begin{aligned}
R_\pm (\rho,z) & = \sqrt{\rho^2 + (z \pm \alpha_+)^2} \quad \Rightarrow  \left. R_\pm \rvert_{z=0} = \sqrt{\rho^2 + \alpha_{+}^2} \equiv R, \\
r_\pm (\rho,z) &  = \sqrt{\rho^2 + (z \pm \alpha_-)^2} \quad \Rightarrow  \left. r_\pm \rvert_{z=0} = \sqrt{\rho^2 + \alpha_{-}^2} \equiv r ,\\ 
\partial_z R_\pm (\rho,z) & = \frac{z \pm \alpha_+}{R_\pm(\rho,z)} \quad \Rightarrow \left. \partial_z R_\pm \rvert_{z = 0} = \frac{\pm \alpha_+}{R} , \\
\partial_z r_\pm (\rho,z) & = \frac{z \pm \alpha_-}{r_\pm(\rho,z)} \quad \Rightarrow \left. \partial_z r_\pm \rvert_{z = 0} = \frac{\pm \alpha_-}{r} .
\end{aligned}
\end{equation}

Let us first take a look at $A$ and its derivative in $z$ at $z = 0$:
\begin{equation}\label{Anotations}
\left. \frac{A(\rho,z)}{\alpha_+ \alpha_- } \rvert_{z = 0} \equiv A_0 \in \mathbb{R}, \qquad \left. \partial_z \frac{A(\rho,z)}{\alpha_+ \alpha_-} \rvert_{z = 0} \equiv \partial_z A_0 \in \mathbb{I}.
\end{equation} 
Similarly, for $B$,
\begin{equation}\label{Bnotations}
\left. \frac{B(\rho,z)}{\alpha_+ \alpha_- } \rvert_{z = 0} \equiv B_0 \in \mathbb{R}, \qquad \left. \partial_z \frac{B(\rho,z)}{\alpha_+ \alpha_-} \rvert_{z = 0} \equiv \partial_z B_0\in \mathbb{I}.
\end{equation} 
Thus we can write the following equalities:
\begin{equation} \label{conjequals}
\begin{aligned}
\bar{A}_0 & = A_0, \qquad \overline{\left(\partial_z A_0 \right)} & = - \partial_z A_0, \\
\bar{A}_0 & = B_0, \qquad \overline{\left(\partial_z B_0 \right)} & = - \partial_z B_0 .
\end{aligned}
\end{equation}

Let us now look at the expression for $f$ and its derivative in $z$:
\begin{equation}
f = \frac{\lvert A \rvert^2 - \lvert B \rvert^2}{\lvert A + B \rvert^2} = \frac{A \overline{A} - B \overline{B}}{(A + B) \overline{(A+B)}}.
\end{equation}
Its derivative in $z$, with $\partial_z A \equiv A'$ and $\partial_z B \equiv B'$ is thus given by:
\begin{equation}
\begin{aligned}
\partial_z f  = \frac{1}{(A+B)^2 (\Ao + \Bo)^2} \Big\{& (A' \Ao + A \Ao' - B' \Bo - B \Bo')(A + B) (\Ao + \Bo)  \\
 & - (A \Ao - B \Bo)\left((A'+ B')(\Ao + \Bo) + (A+B)(\Ao' + \Bo') \right)\Big\}.
\end{aligned}
\end{equation}
When $z =0$, using \eqref{conjequals}, this simplifies to
\begin{equation}
\begin{aligned}
\left. \partial_z f \rvert_{z=0}  = \Bigg( \frac{1}{(A+B)^2 (A+B)^2} \Big\{& (A' A - A A' - B' B + B B')(A + B) (A + B)  \\
 & - (A A - B B)\left((A'+ B')(A + B) - (A+B)(A' + B') \right)\Big\} \Bigg)_{z = 0} \\ 
 & = 0.
\end{aligned}
\end{equation}

For $G$, the situation is a bit more complex, because both the imaginary and the real parts survive when $z = 0$. We will then denote $G$ as $G \equiv G_r + i G_i$.

We can then compute what happens to $G$ and its derivative when $z = 0$.
\begin{equation} \label{Gnotations}
\left. \frac{G}{\alpha_+ \alpha_- } \rvert_{z = 0} \equiv G_r + i G_i, \qquad \left. \partial_z \frac{G}{\alpha_+ \alpha_-} \rvert_{z = 0} \equiv \partial_z G_{r} + i \partial G_{i},
\end{equation}
and in terms of these notations, it follows that:
\begin{equation}
\bar{G}_0  = G_{r} - i G_{i}, \qquad \overline{\left(\partial_z G_0 \right)}  = \partial_z G_r - i \partial_z G_{i}.
\end{equation}

We can now finally compute $\omega$ and its derivative in $z$.
We know that the expression for $\omega$ is given by:
\begin{equation}
\omega(\rho,z) = - 4 \frac{\Im ( G (\overline{A} + \overline{B})) }{\lvert A \rvert^2 - \lvert B \rvert^2},
\end{equation}
thus its derivative in $z$ should look like:
\begin{equation}
\begin{aligned}
\partial_z \omega (\rho,z)  = \frac{-4}{(\lvert A \rvert^2 - \lvert B \rvert^2)^2} \times \Bigg\{ & -4 \Im\left\{ G'(\Ao + \Bo) + G(\Ao' + \Bo')\right\} \left\{ \lvert A \rvert^2 - \lvert B \rvert^2 \right\} \\
& + 4 \Im\left\{ G(\Ao + \Bo)\right\} (A' \Ao + A \Ao' - B' \Bo - B \Bo') \Bigg\}.
\end{aligned}
\end{equation}
We now put $z =0$ and use \eqref{Anotations} and \eqref{Bnotations} to reduce the expression to:
\begin{equation}
\begin{aligned}
\partial_z \omega_0  = \frac{-4}{(\lvert A_0 \rvert^2 - \lvert B_0 \rvert^2)^2} \times \Bigg\{ & -4 \Im\left\{ G_0'(A_0 + B_0) - G_0(A_0'  +B_0')\right\} \left\{ \lvert A_0 \rvert^2 - \lvert B_0 \rvert^2 \right\} \\
& + 4 \Im\left\{ G_0(A_0 + B_0)\right\} (A_0' A_0 - A_0 A_0' - B_0' B_0 + B_0 B_0') \Bigg\}.
\end{aligned}
\end{equation}
The second line vanishes, and we now just have to verify that
\begin{equation}
Q \equiv \Im\left\{ G_0'(A_0 + B_0) - G_0(A_0' + B_0')\right\} = 0.
\end{equation}
To do that, let us use \eqref{Gnotations}, we get
\begin{equation}
Q = \Im\left\{ (G'_{r} + i G'_{i})(A_0 + B_0) - (G_{r} + i G_{i})(A_0' + B_0')\right\}  .
\end{equation}
Since we know that $A_0, B_0 \in \mathbb{R}$ and $A_0', B_0' \in \mathbb{I}$, 
\begin{equation}\label{toverify}
Q =  G'_{i}(A_0 + B_0) - G_{r}(A_0' + B_0'),
\end{equation}
which we will have to brute force our way through.

The easiest way is to realise that 
\begin{equation}
\frac{G'_{i}}{A_0' + B_0'} = m,
\end{equation}
then divide \eqref{toverify} by $(A_0' + B_0')$ and show that $m (A_0 + B_0) - G_{r} = 0$ by looking individually at the factors for $R^2$, $r^2$, $R r$, $R$ and $r$, which are all $0$. Also, there is no independent term.

Thus we find that $Q =0$, which means that $\partial_z \omega_0 = 0$ and the derivatives in \eqref{derivatives} are all identically $0$, which verifies the equation \eqref{cond_z_app} trivially for all $\rho >0, z = 0$.

\printbibliography

\end{document}